\documentclass[5p,times]{elsarticle}

\usepackage{amsmath}            
\usepackage{epstopdf}           
\usepackage{flushend}           

\usepackage{amssymb}

\usepackage{lineno}

\usepackage[figuresright]{rotating}

\usepackage{subfigure}

\begin{document}

\begin{frontmatter}






\title{Optical design for increased interaction length in a high gradient dielectric laser accelerator}


\author[First]{D. Cesar\corref{cor1}}
\ead{dcesar@ucla.edu}

\author[First,Second]{J. Maxson}
\author[First]{P. Musumeci}
\author[First]{X. Shen}

\author[Third]{R.\,J. England}
\author[Third]{K.\,P. Wootton}

\cortext[cor1]{Corresponding Author}

\address[First]{Department of Physics and Astronomy, University of California--Los Angeles, 475 Portola Plaza, Los Angeles, California 90095, USA}
\address[Second]{Present address: Department of Physics, Cornell University, 109 Clark Hall, Ithaca, New York 14853, USA}
\address[Third]{SLAC National Accelerator Laboratory, 2575 Sand Hill Rd, Menlo Park, California 94025, USA}

\begin{abstract}
We present a methodology for designing and measuring pulse front tilt in an ultrafast laser for use in dielectric laser acceleration. Previous research into dielectric laser accelerating modules has focused on measuring high accelerating gradients in novel structures, but has done so only for short electron-laser coupling lengths. Here we demonstrate an optical design to extend the laser-electron interaction to 1\,mm.
\end{abstract}

\end{frontmatter}


\section{Introduction}
Over the last five years there has been a significant effort to develop and validate photonic structures for the direct acceleration of electrons by a laser driver. These devices promise to transform the GV/m fields available from compact commercial laser systems into a modular GV/m linear accelerator\,\cite{england_dielectric_2014}. Towards this end, researchers have design, built, and power-tested a host of innovative photonic structures, demonstrating synchronous laser-electron interaction with fields up to 1.8\,GV/m for relativistic electrons\,\cite{cesar_nonlinear_2017} and up to 370\,MV/m for subrelativistic electrons\,\cite{leedle_dielectric_2015}. Future structure proposals intend to improve field symmetry, damage handling, staging, diagnostics and more\,\cite{mcneur_elements_2016,wei_dual-gratings_2017,kozak_acceleration_2017}

\begin{figure}[]
\centering
\includegraphics[width=.45\textwidth]{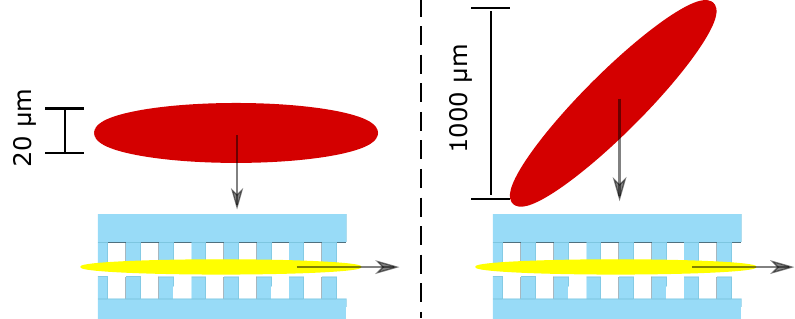}
\caption{Illustration comparing acceleration with a flat pulse (left) to a tilted pulse (right). The flat pulse accelerates a long electron bunch for a short interaction while the tilted pulse accelerates a single slice for an extended interaction.}
\label{fig:EAAC_flat_vs_pft}
\end{figure}

But so far these devices have been operated in a `test' configuration in which the drive laser propagates transversely to the electron beam, limiting the interaction region to the laser pulse-duration. Since the dielectric lase accelerator (DLA) achieves high gradients by using short pulse lengths to avoid material damage, this necessarily limits the particle-wave interaction to a few microns (or limits the gradient \cite{peralta_demonstration_2013}). While there exist designs for fractal-type transverse coupling\,\cite{hughes_-chip_2017} and for longitudinal coupling\,\cite{rosenzweig_galaxie_2012}, experimental demonstrations of extended interactions have yet to be studied. One method to address this challenge is to use free-space optics to impart a pulse-front tilt (PFT) to the incoming laser (Fig.\,\ref{fig:EAAC_flat_vs_pft}) in order to provide ``group velocity matching'' of an ultrafast laser envelope to the electron beam \cite{plettner_proposed_2006}. This technique is similar to that used for phase-matching in the optical rectification of terahertz \cite{hebling_generation_2008}, but the application to electron acceleration provides additional constraints. 


In this proceedings we discuss optical design and diagnostics implemented at the Pegasus photoinjector facility\,\cite{cesar_nonlinear_2017} in order to deliver a pulse-front tilted beam for electron acceleration in a DLA. 

\section{Creating pulse front tilt}
We introduce pulse front tilt by reflecting the drive laser from a single diffraction grating\,\cite{hebling_derivation_1996}. As illustrated in Fig.\,\ref{fig:setup}, the angle of the incident and diffracted beams are not equal so that the length of a ray depends on the transverse beam position, causing the pulse envelope to develop an $x$--$t$ correlation (while the phase fronts remain flat because the grating also adds a phase shift of $2 \pi n_{order}$ per grating period)\,\cite{treacy_optical_1969}. Additional propagation past the grating causes an $x$--$\omega$ correlation and thus an $\omega$--$t$ correlation analogous to that caused by pulse dispersion. To eliminate this dispersion the plane of the diffraction grating\,\cite{kreier_avoiding_2012} can be imaged onto the acceleration plane resulting in a pulse front tilt angle of:
 \begin{equation}\label{eq:pftangle}
 \theta_\textrm{PFT}=\arctan\left(\frac{\lambda_0 m}{d \cos(\theta_d)}\right)
 \end{equation}
where $m$ is the angular magnification of the imaging system, $d$ is the grating pitch, and $\theta_d$ is the diffraction angle which satisfies the condition:
\begin{equation}
d\left(\sin(\theta_i) + \sin(\theta_d)\right) = N \lambda_0.
\end{equation}
 
From this simplified discussion we see the possibility of describing the pulse evolution via propagation of a ray defined by transverse position, transverse angle, longitudinal position (relative to a well-defined pulse), and frequency: $(x, \theta, t, f)$. The canonical example of a pulse which can be characterized by such a ray is the bi-variate Gaussian: $e^{a\left(x\right)^2+2b\left(x t\right)-c\left(t\right)^2 }$ and its Fourier transforms to $(k,\omega)$. Such a formalism has been developed by Kostenbauder\,\cite{kostenbauder_ray-pulse_1990}, used to characterize all first-order spatio-temporal couplings\,\cite{akturk_general_2005}, and extended to the 6D case\,\cite{marcus_spatial_2016}. The utility of this approach will be familiar to accelerator physicists by analogy to the 6D beam transport matrix\,\cite{wiedemann_particle_1999}. We reproduce a few basic properties of the analysis here.

\subsection{Kostenbauder ray matrix formalism}
In the ray-pulse matrix approach a paraxial optical system is reduced to a set of linear relations described by a matrix, $M$:
\begin{equation}\label{eq:rayprop} (x, \theta, t, f)^{T}=M(x, \theta, t, f)^{T}\end{equation}
and propagation of an optical field through the system $M$ is then given by a Huygens kernel \cite{kostenbauder_ray-pulse_1990}:
\begin{equation}\begin{split}\label{eq:huygens} &E\left(x_{out},t_{out}\right)=\eta\int\int dx_{in}dt_{in} E\left(x_{in},t_{in}\right) \times\dots  \\ & \text{exp}\left[-\frac{i \pi}{\lambda_0} \begin{pmatrix} x_{in} \\ x_{out} \\ t_{in}-t_{out}\end{pmatrix}^{T} \begin{pmatrix} \alpha & \beta & \gamma \\ \beta & \delta & \epsilon \\ \gamma & \epsilon & \xi \end{pmatrix} \begin{pmatrix} x_{in} \\ x_{out} \\ t_{in}-t_{out}\end{pmatrix} \right] \end{split}\end{equation}

where the six coefficients $(\alpha,\beta,\gamma,\delta,\epsilon,\xi)$ are given in terms of the matrix $M$. The appearance of combination $(t_{in}-t_{out})$ results from the absence of time-dependent optical elements, while the symmetry of the coefficient matrix is a general property of quadratic forms, so that only six coefficients are needed to describe the optical transport. This suggests that the $4\times4$ matrix $M$ also has only six independent elements: in fact, seven elements are trivial (the output frequency is identical to initial frequency and the input time only affects the output time), while the three nontrivial relations are given by Eq. 15 of Ref.\,\cite{kostenbauder_ray-pulse_1990}. 

For the case of imaging a grating there are only 3 independent parameters and the transfer matrix $M$ is:
\begin{equation}\label{eq:transfermatrix}
M=\begin{pmatrix}
M_{11} & 0 & 0 & 0 \\
M_{21} & \frac{1}{M_{11}} & 0 & \frac{\lambda}{\beta_\textrm{PFT}c}\\
\frac{M_{11}}{\beta_\textrm{PFT}c} & 0 & 1 & 0\\
0 & 0 & 0& 1\\
\end{pmatrix}
\end{equation}
The Huygens kernel for this matrix is a nascent delta function as the imaging condition $M_{12}=0$ is approached so that Eq.\,\ref{eq:huygens} reduces to \cite{kostenbauder_ray-pulse_1990}: 
\begin{equation}\begin{split}
E\left(x_{out},t_{out}\right)&\propto  \int\int dx_{in}dt_{in} \, e^{(-I \pi M_{21} / \lambda_0 M_{11})} \times \dots \\
&\delta(x_{in}-x_{out}/M_{11})  \delta\left(t_{in}-(t_{out}-M_{31}x_{in})\right) E\left(x_{in},t_{in}\right).
\end{split}\end{equation}
When $M_{31}$ is matched to give give the pulse the same velocity as the electrons, the resulting field has the easy interpretation that the electrons travel through the transverse profile of the laser pulse (while remaining stationary in the laser pulse's time dimension). Even outside of perfect imaging the full kernel can be evaluated analytically when the laser is Gaussian. 



\section{Optical design for Pegasus}

\begin{figure}[]
\centering
\includegraphics[width=.45\textwidth]{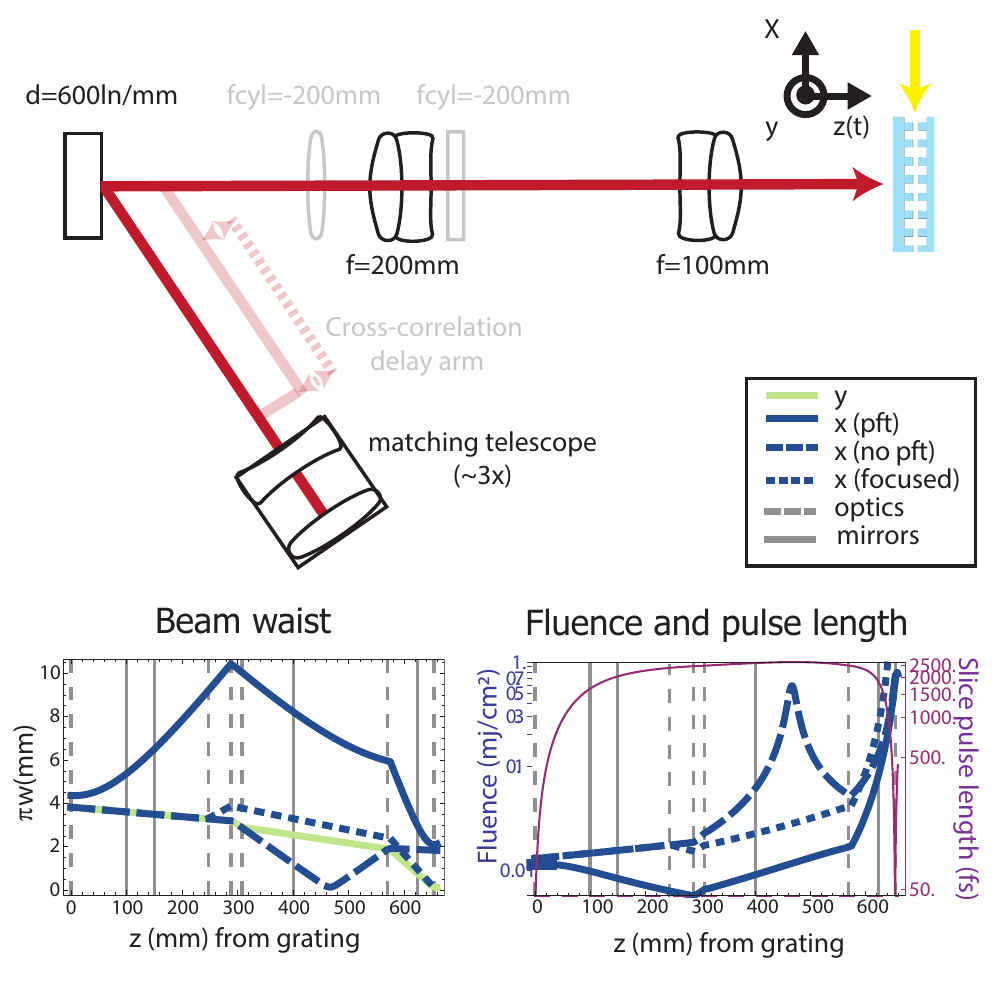}
\caption{Illustration of the optical design. (top) Cartoon showing the relative position of optics. (bottom left) Spot size in 4 operating configurations. (bottom right) Fluence (for 300\,mJ) and slice pulse length (on-axis) used to check damage thresholds.}
\label{fig:setup}
\end{figure}

The minimal optical system required to match a broadband laser for extended interaction at the Pegasus photoinjector test facility is illustrated in Fig.\,\ref{fig:setup} along with the laser waist, beam fluence, and pulse duration. The electron velocity of $\beta=0.997$ (6.5\,MeV) corresponds to a pulse front tilt of $45.1\deg$ which is obtained by using a commercial 600\,ln/mm Au blazed reflection grating and adjusted by tuning the angular magnification of the imaging optics. Demanding independent control of the magnification, image plane, and working distance requires the use of a two-lens telescope, as shown in Fig.\,\ref{fig:setup}.


In addition to setting the pulse front tilt angle, the optics must control the spot-size and phase of the laser at the DLA plane. As observed in previous experiments\,\cite{cesar_nonlinear_2017}, the laser's (slowly varying) wavefront is transfered directly to the field of the accelerating mode such that the laser wavefront will interfere with the particle-wave resonance. Pulse-shaping techniques could be inserted at an intermediate image plane in order to tune the phase of the accelerating mode; however in the absence of pulse-shaping the flat phase requirement means placing the DLA plane at a waist (or in the far field). Control of both the size and radius of curvature requires a two lens matching telescope, which we choose to locate before the diffraction grating. 

Finally, a cylindrical lens is used to focus the in-page dimension of the laser down to a size limited by the electron beam radius ($\sigma_r = 10\,\mu$m). This cylindrical focus allows us to have drastically different fluences at the diffraction grating and at the DLA so that we can take full advantage of the large damage threshold of fused silica\,\cite{soong_laser_2012} compared to the diffraction grating\,\cite{poole_femtosecond_2013}.


In addition to the configurations enumerated above, it is useful, during alignment, to be able to deliver the laser in two alternative configurations: one without the pulse front tilt, and one with a beam tightly focused in both directions in order to maximize fluence. Removing the pulse front tilt, which can be done by replacing the grating with a mirror, simplifies  detection of an acceleration signature by increasing the electron-laser interaction window and reducing the effects of dephasing. Focusing the beam to a small spot, by inserting a second cylindrical lens, is useful for establishing electron-laser time of arrival by burning a copper grid to create an electron plasma which interacts strongly with the main electron beam\,\cite{scoby_effect_2013}.



\subsection{Aberration and wavefront tilt}
While the preceding discussion has focused on the linear optical design, a number of nonlinear aberrations may be considered. Firstly, achromatic lenses should be used in the imaging system or else individual frequency components can not be simultaneously imaged. This has the effect of temporally broadening the pulse by creating a curved pulse front at the image plane\,\cite{bor_distortion_1988}. Secondly, the diffraction grating must lie in the same plane as the imaging lenses (i.e. both lenses perpendicular to the pulse propagation) in order to prevent temporal broadening in the wings of the pulse \cite{kreier_avoiding_2012}. The remaining aberrations (spherical etc.) can be studied by ray-tracing and are found to be tolerable provided the laser is not too large compared to the focal length of the lens.

\section{Measuring pulse front tilt}
\begin{figure}[]
\centering
\includegraphics[width=0.4\textwidth]{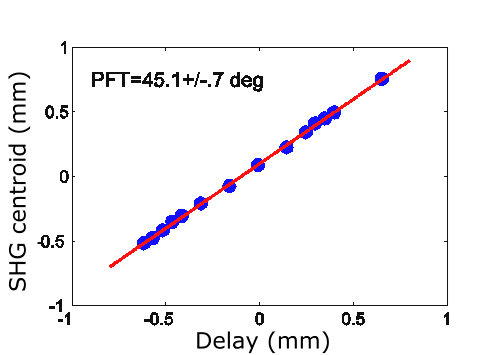}
\caption{Pulse front tilt measurement from the cross-correlation of a flat and tilted pulse. At each delay the flat pulse overlaps with the tilted pulse in small region of $x$--$t$ creating an shg signal centered on the location of the blue data. The red line is a linear fit whose slope is $-\beta_{pft}$}
\label{fig:crosscorrelation}
\end{figure}

Characterization of the pulse front tilt angle is obtained by cross-correlating the tilted pulse with a replica of the original flat pulse \cite{kreier_avoiding_2012}, as illustrated by the opaque delay line in Fig.\,\ref{fig:setup}. As the delay is scanned, the flat pulse pulse overlaps with different temporal slices of the tilted pulse, which, because of the tilt, correspond to different transverse slices. To measure the intensity profile we replace the DLA with a thin (100\,$\mu$m) type-II second harmonic generation (shg) $\alpha$-beta barium borate (BBO) crystal and use polarizing beam-splitters for splitting or re-combining the pulses. For a pulse of (magnified) size $M_{11}w_0$, pulse length $\Delta t_0$ and pulse front tilt ``velocity'' $\beta_\textrm{PFT}$ at the BBO, the shg signal will be a function of the transverse position ($x$) and the delay between the cross-correlating pulses ($\delta_t$):  

\begin{equation}\label{eq:crosscorrelationsignal}
\begin{split}
I &\propto \chi^{(2)}_{x,y,x}\left(2\omega,\omega,\omega\right)E_{x}\left(\omega\right)E_{y}\left(\omega\right) \\
&\propto \exp\left(-2\left(\frac{x}{M_{11}w_0}\right)^2-\frac{1}{2}\left(\frac{x+\beta_\textrm{PFT}c\delta t}{\beta_\textrm{PFT}c\Delta t_0}\right)^2\right)
\end{split}
\end{equation}
where the first term gives the laser spot size and second gives the cross-correlation term. $M_{11}w_0>>\beta_\textrm{PFT}c\Delta t_0$ so that measuring the intensity centroid (in $x$) vs delay ($\delta t$), as shown in Fig.\,\ref{fig:crosscorrelation} gives a measurement of the line ($x+\beta_\textrm{PFT}c\delta t$), allowing us to determine the pulse front tilt between the two pulses.

To obtain an unbiased measurement of the pulse front tilt angle it is important to carefully co-align the interfering pulses in the near and far-field (or alternatively to reduce interferometric fringes in type-I SHG) since any angle between the pulses will appear as additional pulse front tilt. Having achieved alignment of better than $0.5\deg$ our remaining  precision is set by the centroid resolution (the larger of $\beta_\textrm{PFT}c\Delta t_0$ and the camera resolution) and the length over which deviations can be measured ($M_{11}w_0$). For our measurements this statistical uncertainty is typically $<0.4\deg$, for an overall uncertainty of less than $0.7\deg$.

\section{Alignment and distortions}
Extending the interaction length to thousands of optical periods puts strict tolerances on the electron-laser alignment. While the group velocity matching is a robust procedure  ($d\theta_\textrm{PFT}/d\theta_{in}$=0) which can be checked by the cross-correlation diagnostic, co-aligning the narrow laser envelope ($1\,\textrm{mm}\times\,45\,{\mu}\textrm{m}$) and pulse front direction to the electron beam path must be done by carefully referencing each path to an external marker (e.g. alignment channels in the DLA). Furthermore, the DLA's sensitivity to changes in the phase of the drive laser\,\cite{cesar_nonlinear_2017} increases with interaction length, which makes control of the wavefront an important tool for studying beam dynamics.

\subsection{Wavefront distortions}
By far the most important wavefront distortion is wavefront tilt. For a plane wave $(\omega,k_\perp=k_0\sin(\theta))$ to interact with an electron it must satisfy: $k_g z+k_0\sin\left(\theta\right)z-\omega(\theta) \left(z/\beta c\right)=0$, where the angular dispersion $\omega(\theta)$ is what allows an electron to interact with multiple frequencies. The wavefront tilt, $\theta$, must be precisely set by aligning the back-reflection (of a known frequency component) off of the DLA and then fine tuned ($<1$\,mrad) by tilting the DLA to control the accelerating mode velocity. 

Of the higher order wavefront errors, we can eliminate the quadratic term by design, but the others may remain. The third (and higher) order aberrations, however, only cause dephasing in the wings of the pulse, where the intensity is low. Based on a wavefront measurement between the matching telescope and the grating we can calculate that third and higher order aberrations will cause $<5$\% reduction in the electron energy gain during experiments at Pegasus.

\subsection{Kerr nonlinearity}
At high intensities the glass substrate of the DLA acts as a nonlinear Kerr lens, disrupting the acceleration process by adding an intensity dependent phase $\Delta \Phi = n_2 k_0 L I$ to the electric field. We could compensate for this phase by using a radius of curvature to offset the phase, analogous to the experiment in the time-domain\,\cite{cesar_nonlinear_2017}; i.e. placing the low-intensity waist before the DLA and allowing the beam to self-focus. Given the transfer matrix (Eq.\,\ref{eq:transfermatrix}) from the grating to the DLA and setting the radius of curvature at the DLA equal to second order curvature of a Gaussian beam ($\Delta \Phi / w_{dla}^2$) we get an approximate condition for compensation:
\begin{equation}
\frac{k_0 \left(M_{11} M_{21}R_{in}+1\right)}{2 M_{11}^2 R_{in}}=\frac{\Delta \Phi}{M_{11}^2 w_{in}^2}
\end{equation}
For a fixed imaging system certain values of $\Delta \Phi$ can require an unreasonable input beam, however by varying $M_{11}$,or by adding a third lens to separately control $M_{21}$, the compensation can always be arranged. At any rate, the problem can be mitigated by using a larger $w_\textrm{DLA}$ so that the intensity is roughly constant over the accelerating structure.

\begin{figure}[]
\centering
\includegraphics[width=.45\textwidth]{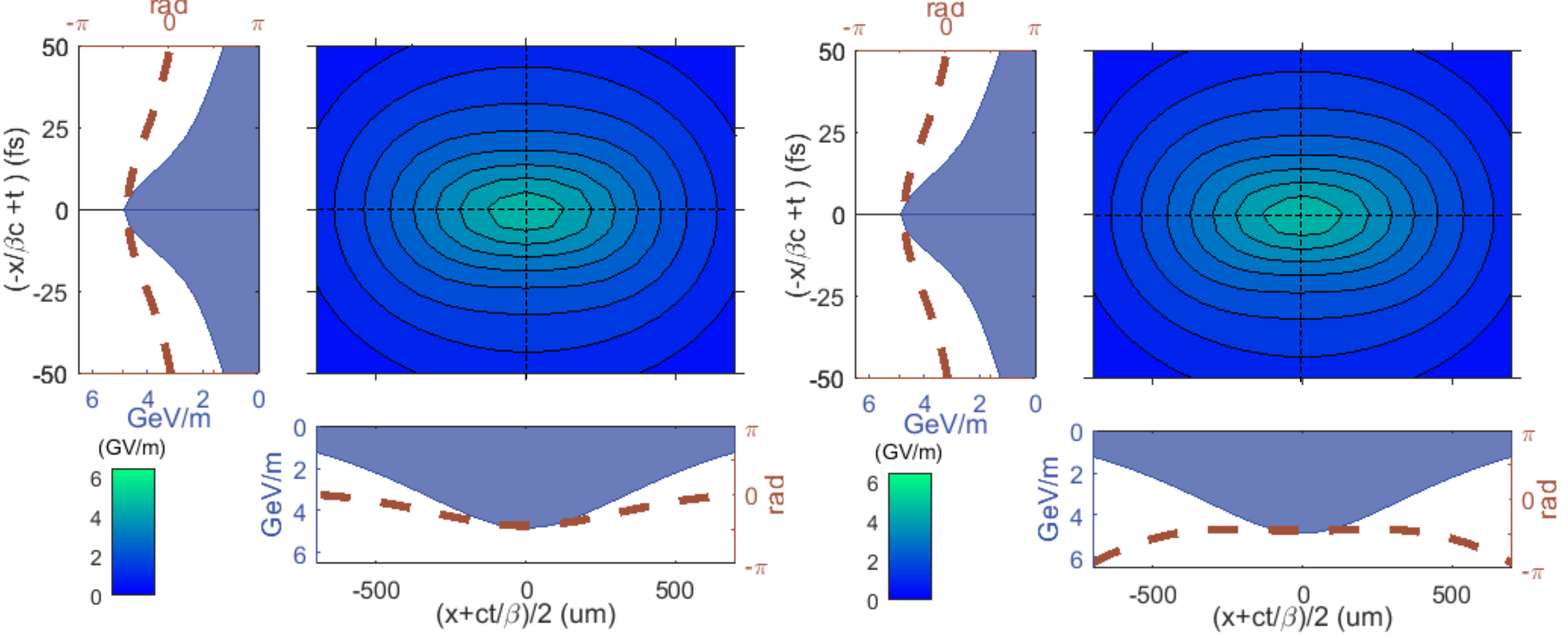}
\caption{Laser profiles after nonlinear propagation. (left) The uncorrected, distorted pulse. (right) The phase-compensated pulse. The contours show (x-t) profiles of $|E|$ demonstrating self-focusing, and lineouts (along the dotted lines) show the amplitude and phase on-axis. Electrons propagate along the horizontal axis so that comparing the bottom lineouts show how the compensated pulse flattens the accelerating phase.}
\label{fig:nonlin}
\end{figure}

This heuristic model works well in the time-domain, where to first order the nonlinearity adds a phase but leaves the pulse shape unchanged, but it is more complicated in the spatial domain where self-focusing steepens the pulse profile. The self-focusing becomes even more pronounced in a pulse-front tilted laser since at each ``$t$'' slice of the beam the ``$x$'' profile is much narrower than in the equivalent flat pulse. To model this distortion we adapt the split-step code used in \cite{cesar_nonlinear_2017} for pulse-front-tilt by re-writing the propagator in a coordinate system $(- x/\beta c + t, x+ c t/\beta)$ directed along the axes of the PFT envelope (which are orthogonal in units where $c=1$). 

The resulting field profiles are shown in Fig.\,\ref{fig:nonlin} where the self-focusing is evident in the oval shape of the contours in the high intensity region (compared to circular contours at low intensity). The self-focusing is primarily in the time-like coordinate, in which the electron trajectories are stationary, so that its primary effect is to raise $I$ and thus $\Delta \Phi$ (from $\pi/3$ to $\pi/2$ in this case). Despite this, we can still change the radius of curvature input to the optical system and compensate for the nonlinear phase. This is demonstrated in the right half of Fig.\,\ref{fig:nonlin} where the phase along an electron trajectory (horizontal lineout) is flat in the high-intensity region.

In addition to altering $\Delta \Phi$ the self-focusing will change the energy distribution of the accelerated electron bunch (assuming the electron bunch is longer than the pulse duration). For a Gaussian laser pulse many electrons experience the maximum gradient, leading to a sharp edge in the electron spectra; while after self-focusing the laser approaches a Townes profile which will make an electron spectra with a long high energy tail.


\section{Conclusion}
We have described the design and characterization of a free space optical system for studying extended interactions in DLA. This experimental technique should prove valuable for a number of proposed experiments \cite{prat_outline_2017,wei_dual-grating_2017} intending to use a tilted pulse front. By application to DLA tests at the Pegasus facility we show how pulse distortions may be controlled in order to optimize acceleration and allow for beam dynamics to dominate the interaction. 

\section*{Acknowledgements}
This work was supported by the Gordon and Betty Moore Foundation under grant GBMF4744 (Accelerator on a Chip) and by the U.S. Department of Energy under Contract DE-AC02-76SF00515.

\label{}





\bibliographystyle{elsarticle-num}
\bibliography{EAAC_References.bib}

\end{document}